\documentclass{article}
\usepackage[T1]{fontenc}
\usepackage{a4wide}
\makeatletter

\newcommand{\LyX}{L\kern-.1667em\lower.25em\hbox{Y}\kern-.125emX\@}

\newcommand{\lyxaddress}[1]{
  \par {\raggedright #1 
  \vspace{1.4em}
  \noindent\par}
}

\makeatother

\begin{document}

\title{\noindent Entanglement Teleportation Through Cat-like States}

\author{Sibasish Ghosh\protect\( ^{1}\protect \)\thanks{
res9603@isical.ac.in
}, Guruprasad Kar\protect\( ^{1}\protect \), Anirban Roy\protect\( ^{1}\protect \)\thanks{
res9708@isical.ac.in
}, Debasis Sarkar\protect\( ^{2}\protect \) and Ujjwal Sen\protect\( ^{3}\protect \)\thanks{
dhom@boseinst.ernet.in
}}

\maketitle

\lyxaddress{\protect\( ^{1}\protect \)Physics and Applied Mathematics Unit, Indian Statistical
Institute, 203 B. T. Road, Calcutta 700035, India }

\lyxaddress{\protect\( ^{2}\protect \)Department of Mathematics, Burdwan University,
Burdwan, West Bengal, India }

\lyxaddress{\protect\( ^{3}\protect \)Department of Physics, Bose Institute, 93/1 A.
P. C. Road, Calcutta 700009, India}

\begin{abstract}
\noindent{We first consider teleportation of entangled states shared between
Claire and Alice to Bob1 and Bob2 when Alice and the two Bobs share a single
copy of a GHZ-class state and where \emph{all} the four parties are at distant
locations. We then generalize this situation to the case of teleportation of
entangled states shared between Claire1, Claire2,....., Claire(N-1) and Alice
to Bob1, Bob2,....., BobN when Alice and the N Bobs share a single copy of a
Cat-like state and where again \emph{all} the 2N parties are at distant locations.} 
\end{abstract}
\noindent Quantum teleportation, proposed by Bennett \emph{et al.}(BBCJPW)\cite{1},
is a protocol by which an arbitrary qubit can be transferred (teleported) exactly
from one location (where say, Alice is operating) to a possibly distant location
(operated say, by Bob) by using only local operations and classical communication,
without sending the particle itself. This almost impossibility is made possible
by allowing Alice and Bob to \emph{a priori} share a maximally entangled state
of two qubits say, 
\[
\left| \phi ^{+}\right\rangle =\frac{1}{\sqrt{2}}(\left| 00\right\rangle +\left| 11\right\rangle )\]
It is important to note that the entanglement in the channel vanishes completely
after it has been used to send the qubit by using the BBCJPW protocol. Now if
the sent qubit is \emph{a priori} entangled with another qubit (possessed by
Claire), it will remain so after teleportation. That is, the previously shared
entanglement between Alice and Claire would now be shared between Bob and Claire. 

\noindent Consider now a different situation. A source delivers an arbitrary
two-qubit entangled state to Alice which must finally be shared between Bob1
and Bob2. Instead of state teleportation, Alice therefore has the task of entanglement
teleportation. It would be sufficient if Alice shares a maximally entangled
state with Bob1 and another with Bob2. Alice would then just teleport the two
qubits using the BBCJPW protocol.\cite{2,3}

\noindent But what if Alice shares with the Bob1-Bob2 system, less than two
ebits of entanglement? Suppose for example that instead of the two maximally
entangled states, Alice, Bob1 and Bob2 share the GHZ state\cite{4} 
\[
\left| GHZ\right\rangle =\frac{1}{\sqrt{2}}(\left| 000\right\rangle +\left| 111\right\rangle )\]
Would the same feat be possible now? Gorbachev and Trubilko\cite{5} have considered
this case and shown that if Alice knows that the state has been prepared in
the Schmidt basis \( \{\left| 0^{\prime }0^{\prime \prime }\right\rangle ,\: \left| 1^{\prime }1^{\prime \prime }\right\rangle \} \),
i.e., if Alice knows that the state is of the form
\[
\left| \chi \right\rangle =\alpha \left| 0^{\prime }0^{\prime \prime }\right\rangle +\beta \left| 1^{\prime }1^{\prime \prime }\right\rangle \]
with known \( \left| 0^{\prime }0^{\prime \prime }\right\rangle  \) and \( \left| 1^{\prime }1^{\prime \prime }\right\rangle  \)
but unknown Schmidt coefficients \( \alpha  \), \( \beta  \), then this state
can be made to share between Bob1 and Bob2. In this paper we simplify their
protocol. Shi \emph{et al.}\cite{6} generalized this situation to the case
in which the state \( \left| \chi \right\rangle  \) \emph{}is shared between
\emph{two separated parties} Alice and Claire.\cite{7} The protocol of Shi
\emph{et al.} is probabilistic as in their case Alice and the two Bobs share
the state 
\[
\left| GHZ^{\prime }\right\rangle =a\left| 000\right\rangle +b\left| 111\right\rangle \]
instead of a GHZ. Here we show that even if Alice and the two Bobs share the
state
\[
\left| ghz\right\rangle =\frac{1}{\sqrt{2}}(\left| 0\phi 0\right\rangle +\left| 1\phi ^{\prime }1\right\rangle )\]
where \emph{\( \left| \phi \right\rangle  \) and \( \left| \phi ^{\prime }\right\rangle  \)
are not necessarily orthogonal}, it is possible for Alice and Claire to make
the two Bobs share the state
\[
\left| \chi ^{\prime }\right\rangle =\alpha \left| \phi 0^{\prime }\right\rangle +\beta \left| \phi ^{\prime }1^{\prime }\right\rangle \]
 Our protocol is independent of the ones in ref.\cite{8} and much simpler.
More important is the fact that our protocol is generalizable to the N-party
situation. We also touch the probabilistic case in both situations. 

Suppose a source prepares the state \( \left| \chi \right\rangle  \) (with
known \( \left| 0^{\prime }0^{\prime \prime }\right\rangle  \) and \( \left| 1^{\prime }1^{\prime \prime }\right\rangle  \)
but unknown \( \alpha  \), \( \beta  \)) and delivers it to Alice who wants
to make it shared between Bob1 and Bob2 through a GHZ state which she shares
with the Bobs. This situation has been considered in ref.\cite{5}. We simplify
their protocol and show that \( \left| \chi \right\rangle  \) can be made to
share between the two Bobs by simply using the BBCJPW protocol. Indeed \( \left| \chi \right\rangle  \)
is essentially a qubit. What is important is that there is no nonlocal operation
involved between the two Bobs in the protocol. Let us elaborate.

First Alice transforms \( \left| \chi \right\rangle  \) to
\[
\left| \xi \right\rangle =\alpha \left| 00\right\rangle +\beta \left| 11\right\rangle \]
which is possible as the Schmidt basis \( \{\left| 0^{\prime }0^{\prime \prime }\right\rangle ,\: \left| 1^{\prime }1^{\prime \prime }\right\rangle \} \)
is known. The combined state \( \left| \xi \right\rangle _{12}\left| GHZ\right\rangle _{AB_{1}B_{2}} \)
may be written as
\[
\frac{1}{2}[\left| \phi _{GHZ}^{+}\right\rangle _{12A}\otimes (\alpha \left| 00\right\rangle +\beta \left| 11\right\rangle )_{B_{1}B_{2}}+\left| \phi _{GHZ}^{-}\right\rangle _{12A}\otimes (\alpha \left| 00\right\rangle -\beta \left| 11\right\rangle )_{B_{1}B_{2}}\]
\[
+\left| \psi _{GHZ}^{+}\right\rangle _{12A}\otimes (\alpha \left| 11\right\rangle +\beta \left| 00\right\rangle )_{B_{1}B_{2}}+\left| \psi _{GHZ}^{-}\right\rangle _{12A}\otimes (\alpha \left| 11\right\rangle -\beta \left| 00\right\rangle )_{B_{1}B_{2}}]\]
where
\[
\left| \phi _{GHZ}^{\pm }\right\rangle =\frac{1}{\sqrt{2}}(\left| 000\right\rangle \pm \left| 111\right\rangle )\]
 
\[
\left| \psi _{GHZ}^{\pm }\right\rangle =\frac{1}{\sqrt{2}}(\left| 001\right\rangle \pm \left| 110\right\rangle )\]
Alice now performs a projection-valued measurement on her three qubits (1, 2
and A) with the projection operators 
\[
P^{GHZ}_{1}=P\left[ \left| \phi _{GHZ}^{+}\right\rangle \right] ,\: P^{GHZ}_{2}=P\left[ \left| \phi _{GHZ}^{-}\right\rangle \right] \]
\[
P^{GHZ}_{3}=P\left[ \left| \psi _{GHZ}^{+}\right\rangle \right] ,\: P^{GHZ}_{4}=P\left[ \left| \psi _{GHZ}^{-}\right\rangle \right] \]
\[
P_{5}^{GHZ}=I-\sum ^{4}_{i=1}P^{GHZ}_{i}\]
and communicates the result to Bob1 and Bob2. If \( P^{GHZ}_{1} \) clicks,
the Bobs are to do nothing. They would then already share the state \( \left| \xi \right\rangle  \).
If \( P^{GHZ}_{2} \) clicks, Bob2 does nothing but Bob1 applies \( \sigma _{z} \)
on his particle. If \( P^{GHZ}_{3} \) clicks, both of them apply \( \sigma _{x} \)
i.e., they apply \( \sigma _{x}\otimes \sigma _{x} \). And if \( P^{GHZ}_{4} \)
clicks they apply \( \sigma _{x}\otimes i\sigma _{y} \). \( P^{GHZ}_{5} \)
would never click. Finally, Bob1 and Bob2 share the state \( \left| \xi \right\rangle  \)
on which they apply \( U_{1}\otimes U_{2} \) to transform it to \( \left| \chi \right\rangle  \)
where \( U_{1} \) is the unitary operator that transforms \( \left| 0\right\rangle \rightarrow \left| 0^{\prime }\right\rangle  \)
and \( \left| 1\right\rangle \rightarrow \left| 1^{\prime }\right\rangle  \)
and \( U_{2} \) the unitary operator that transforms \( \left| 0\right\rangle \rightarrow \left| 0^{\prime \prime }\right\rangle  \)
and \( \left| 1\right\rangle \rightarrow \left| 1^{\prime \prime }\right\rangle  \).
Note that throughout the process there is no nonlocal operation involved between
Bob1 and Bob2. 

Shi \emph{et al.}\cite{6} have generalized this situation to the case in which
two separated parties Alice and Claire share the state \( \left| \chi \right\rangle  \).
We show that even in the case in which Alice and the two Bobs share the state
\[
\left| ghz\right\rangle =\frac{1}{\sqrt{2}}(\left| 0\phi 0\right\rangle +\left| 1\phi ^{\prime }1\right\rangle )\]
where \( \left| \phi \right\rangle  \) and \( \left| \phi ^{\prime }\right\rangle  \)
are not necessarily orthogonal, it is possible for Alice and Claire to make
the two Bobs share the state 
\[
\left| \chi ^{\prime }\right\rangle =\alpha \left| \phi 0^{\prime }\right\rangle +\beta \left| \phi ^{\prime }1^{\prime }\right\rangle \]
Note that \( \left| ghz\right\rangle  \) is a state of the ``GHZ-class''\cite{8}. 

The initial combined state is 
\[
\left| \chi ^{\prime }\right\rangle _{12}\left| ghz\right\rangle _{AB_{1}B_{2}}=\left( \alpha \left| \phi 0^{\prime }\right\rangle +\beta \left| \phi ^{\prime }1^{\prime }\right\rangle \right) _{12}\frac{1}{\sqrt{2}}\left( \left| 0\phi 0\right\rangle +\left| 1\phi ^{\prime }1\right\rangle \right) _{AB_{1}B_{2}}\]
 where the particles 1 and 2 belong to Claire and Alice respectively. First
of all Alice performs the unitary operation \( U^{\prime } \), on the qubit
2, that transforms \( \left| 0^{\prime }\right\rangle \rightarrow \left| 0\right\rangle  \)
and \( \left| 1^{\prime }\right\rangle \rightarrow e^{-i\varepsilon }\left| 1\right\rangle  \)
where \( \left\langle \phi \right. \left| \phi ^{\prime }\right\rangle =re^{i\varepsilon } \)
so that \( \left| \chi ^{\prime }\right\rangle _{12} \) transforms to
\[
\left| \xi ^{\prime }\right\rangle _{12}=\alpha \left| \phi 0\right\rangle +\beta \left| \phi ^{\prime \prime }1\right\rangle \]
where \( \left| \phi ^{\prime \prime }\right\rangle =e^{-i\varepsilon }\left| \phi ^{\prime }\right\rangle  \).
Alice also applies the unitary operator, on qubit A, that transforms \( \left| 0\right\rangle \rightarrow \left| 0\right\rangle  \)
and \( \left| 1\right\rangle \rightarrow e^{-i\varepsilon }\left| 1\right\rangle  \)
so that \( \left| ghz\right\rangle  \) transforms to
\[
\left| ghz_{1}\right\rangle =\frac{1}{\sqrt{2}}(\left| 0\phi 0\right\rangle +\left| 1\phi ^{\prime \prime }1\right\rangle )\]
 The state of the five particles is now
\[
\left| \xi ^{\prime }\right\rangle _{12}\left| ghz_{1}\right\rangle _{AB_{1}B_{2}}=(\alpha \left| \phi 0\right\rangle +\beta \left| \phi ^{\prime }1\right\rangle )\frac{1}{\sqrt{2}}(\left| 0\phi 0\right\rangle +\left| 1\phi ^{\prime \prime }1\right\rangle )\]
which may be rewritten as 
\[
\frac{1}{2}[\left( \alpha \left| \phi \phi 0\right\rangle +\beta \left| \phi ^{\prime \prime }\phi ^{\prime \prime }1\right\rangle \right) _{1B_{1}B_{2}}\otimes \left| \phi ^{+}\right\rangle _{2A}+\left( \alpha \left| \phi \phi 0\right\rangle -\beta \left| \phi ^{\prime \prime }\phi ^{\prime \prime }1\right\rangle \right) _{1B_{1}B_{2}}\otimes \left| \phi ^{-}\right\rangle _{2A}\]
 
\[
+\left( \alpha \left| \phi \phi ^{\prime \prime }1\right\rangle +\beta \left| \phi ^{\prime \prime }\phi 0\right\rangle \right) _{1B_{1}B_{2}}\otimes \left| \psi ^{+}\right\rangle _{2A}+\left( \alpha \left| \phi \phi ^{\prime \prime }1\right\rangle -\beta \left| \phi ^{\prime \prime }\phi 0\right\rangle \right) _{1B_{1}B_{2}}\otimes \left| \psi ^{-}\right\rangle _{2A}]\]
 where
\[
\left| \phi ^{\pm }\right\rangle =\frac{1}{\sqrt{2}}(\left| 00\right\rangle \pm \left| 11\right\rangle )\]
 
\[
\left| \psi ^{\pm }\right\rangle =\frac{1}{\sqrt{2}}(\left| 01\right\rangle \pm \left| 10\right\rangle )\]
Alice now conducts a projection measurement (the Bell measurement) on her two
particles with the projection operators
\[
P_{1}=P[\left| \phi ^{+}\right\rangle ],\: P_{2}=P[\left| \phi ^{-}\right\rangle ]\]
\[
P_{3}=P[\left| \psi ^{+}\right\rangle ],\: P_{4}=P[\left| \psi ^{-}\right\rangle ]\]
 After that she sends two bits of classical message to each of Bob1 and Bob2
to tell them the result of the Bell measurement. 

As \( \left\langle \phi \right. \left| \phi ^{\prime \prime }\right\rangle \: (=r) \)
belongs to \( [0,\: 1] \), there exists a unique orthonormal basis \( \{\left| a\right\rangle ,\: \left| \overline{a}\right\rangle \} \)
such that
\[
\left| \phi \right\rangle =\cos \frac{\theta }{2}\left| a\right\rangle +\sin \frac{\theta }{2}\left| \overline{a}\right\rangle \]
\[
\left| \phi ^{\prime }\right\rangle =\cos \frac{\theta }{2}\left| a\right\rangle -\sin \frac{\theta }{2}\left| \overline{a}\right\rangle \]
where \( \theta \: \epsilon \: [0,\: \pi /2] \). Note that \( \theta  \),
\( \left| a\right\rangle  \), \( \left| \overline{a}\right\rangle  \) are
all known. Claire performs a projective measurement on just the basis \( \{\left| a\right\rangle ,\: \left| \overline{a}\right\rangle \} \)
and communicates the result to the Bobs. 

It is now straightforward to see that there always exists a \emph{product}-unitary
operation between the two Bobs, depending upon the results communicated by Alice
and Claire, so that they (the Bobs) end up sharing the state \( \left| \xi ^{\prime }\right\rangle  \).
If \( P_{1} \) clicks in Alice's measurement, then Claire and the two Bobs
share the state 
\[
\alpha \left| \phi \phi 0\right\rangle _{1B_{1}B_{2}}+\beta \left| \phi ^{\prime \prime }\phi ^{\prime \prime }1\right\rangle _{1B_{1}B_{2}}\]
\[
=\alpha \cos \frac{\theta }{2}\left| a\phi 0\right\rangle +\alpha \sin \frac{\theta }{2}\left| \overline{a}\phi 0\right\rangle +\beta \cos \frac{\theta }{2}\left| a\phi ^{\prime \prime }1\right\rangle -\beta \sin \frac{\theta }{2}\left| \overline{a}\phi ^{\prime \prime }1\right\rangle \]
\[
=\cos \frac{\theta }{2}\left| a\right\rangle _{1}\left( \alpha \left| \phi 0\right\rangle +\beta \left| \phi ^{\prime \prime }1\right\rangle \right) _{B_{1}B_{2}}+\sin \frac{\theta }{2}\left| \overline{a}\right\rangle _{1}\left( \alpha \left| \phi 0\right\rangle -\beta \left| \phi ^{\prime \prime }1\right\rangle \right) _{B_{1}B_{2}}\]
 If after her measurement, Claire obtains the result \( \left| a\right\rangle  \),
then the Bobs are to do nothing. On the other hand if she obtains the result
\( \left| \overline{a}\right\rangle  \), only Bob2 is to perform a unitary
operation that transforms \( \left| 0\right\rangle \rightarrow \left| 0\right\rangle  \)
and \( \left| 1\right\rangle \rightarrow -\left| 1\right\rangle  \), which
is just \( \sigma _{z} \). 

If \( P_{2} \) clicks, then Claire and the Bobs share the state 
\[
\alpha \left| \phi \phi 0\right\rangle _{1B_{1}B_{2}}-\beta \left| \phi ^{\prime \prime }\phi ^{\prime \prime }1\right\rangle _{1B_{1}B_{2}}\]
\[
=\cos \frac{\theta }{2}\left| a\right\rangle _{1}\left( \alpha \left| \phi 0\right\rangle -\beta \left| \phi ^{\prime \prime }1\right\rangle \right) _{B_{1}B_{2}}+\sin \frac{\theta }{2}\left| \overline{a}\right\rangle _{1}\left( \alpha \left| \phi 0\right\rangle +\beta \left| \phi ^{\prime \prime }1\right\rangle \right) _{B_{1}B_{2}}\]
 In this case, the Bobs are to do nothing if Claire obtains \( \left| \overline{a}\right\rangle  \)
and only Bob2 is to apply \( \sigma _{z} \) if Claire obtains \( \left| a\right\rangle  \). 

If \( P_{3} \) clicks, then the shared state is
\[
\alpha \left| \phi \phi ^{\prime \prime }1\right\rangle _{1B_{1}B_{2}}+\beta \left| \phi ^{\prime \prime }\phi 0\right\rangle _{1B_{1}B_{2}}\]
\[
=\cos \frac{\theta }{2}\left| a\right\rangle _{1}\left( \alpha \left| \phi ^{\prime \prime }0\right\rangle +\beta \left| \phi 1\right\rangle \right) _{B_{1}B_{2}}+\sin \frac{\theta }{2}\left| \overline{a}\right\rangle _{1}\left( \alpha \left| \phi ^{\prime \prime }0\right\rangle -\beta \left| \phi 1\right\rangle \right) _{B_{1}B_{2}}\]
 Since the inner product of \( \left| \phi \right\rangle  \) and \( \left| \phi ^{\prime \prime }\right\rangle  \)
is real, there exists a unitary operator \( U^{\prime \prime } \) that transforms
\( \left| \phi \right\rangle \rightarrow \left| \phi ^{\prime \prime }\right\rangle  \)
and \( \left| \phi ^{\prime \prime }\right\rangle \rightarrow \left| \phi \right\rangle  \).
Irrespective of what Claire obtained, Bob1 is to apply just this operator. However
Bob2 is to do nothing if Claire obtains \( \left| a\right\rangle  \) but to
apply \( \sigma _{z} \) if Claire obtains \( \left| \overline{a}\right\rangle  \). 

If \( P_{4} \) clicks in Alice's measurement, Bob1 again applies \( U^{\prime \prime } \)
irrespective of Claire's result. And Bob2 is to do nothing if Claire obtains
\( \left| \overline{a}\right\rangle  \) and to apply \( \sigma _{z} \) if
she obtains \( \left| a\right\rangle  \). At the end of all these the Bobs
are left with the state \( \left| \xi ^{\prime }\right\rangle  \) on which
Bob2 has to perform a rotation to transform it to \( \left| \chi ^{\prime }\right\rangle  \).
Precisely Bob2 has to apply \( \left( U^{\prime }\right) ^{-1} \).

Looking at the protocol described above, which is essentially a generalization
of the BBCJPW protocol, it may seem that the measurement of Claire must be preceded
by that of Alice, and so Alice must communicate to Claire that her (Alice's)
measurement has been performed. But that is not true. The protocol would go
through irrespective of whether Alice or Claire performed the first measurement.
Indeed the measurements are to be performed in two different Hilbert spaces
and the corresponding projection operators would therefore commute. For example
if Alice obtains \( \left| \phi ^{+}\right\rangle  \) and Claire obtains \( \left| a\right\rangle  \),
the Bobs would share the state \( \alpha \left| \phi 0\right\rangle +\beta \left| \phi ^{\prime \prime }1\right\rangle  \)
irrespective of who performed the first measurement. 

For completeness, note that if Alice and two Bobs share the state 
\[
\left| ghz^{\prime }\right\rangle =a\left| 0\phi 0\right\rangle +b\left| 1\phi ^{\prime }1\right\rangle \]
the above entanglement teleportation is possible in a probabilistic manner where
Alice has to change her operations in the same way as exact teleportation was
changed to probabilistic teleportation in ref.\cite{9}. Suffice it to mention
that the combined state 
\[
\left| \xi ^{\prime }\right\rangle _{12}\left| ghz_{1}^{\prime }\right\rangle _{AB_{1}B_{2}}=(\alpha \left| \phi 0\right\rangle +\beta \left| \phi ^{\prime \prime }1\right\rangle )(a\left| 0\phi 0\right\rangle +b\left| 1\phi ^{\prime \prime }1\right\rangle )\]
may be written as
\[
\frac{1}{2}\{(\alpha \left| \phi \phi 0\right\rangle +\beta \left| \phi ^{\prime \prime }\phi ^{\prime \prime }1\right\rangle )_{1B_{1}B_{2}}\otimes (a\left| 00\right\rangle +b\left| 11\right\rangle )_{2A}\]
\[
+(\alpha \left| \phi \phi 0\right\rangle -\beta \left| \phi ^{\prime \prime }\phi ^{\prime \prime }1\right\rangle )_{1B_{1}B_{2}}\otimes (a\left| 00\right\rangle -b\left| 11\right\rangle )_{2A}\]
 
\[
+(\alpha \left| \phi \phi ^{\prime \prime }1\right\rangle +\beta \left| \phi ^{\prime \prime }\phi 0\right\rangle )_{1B_{1}B_{2}}\otimes (a\left| 01\right\rangle +b\left| 10\right\rangle )_{2A}\]
\[
+(\alpha \left| \phi \phi ^{\prime \prime }1\right\rangle -\beta \left| \phi ^{\prime \prime }\phi 0\right\rangle )_{1B_{1}B_{2}}\otimes (a\left| 01\right\rangle -b\left| 10\right\rangle )_{2A}\}\]

We now go over to the N-party case. Suppose there are N Bobs, Bob1, Bob2,.....,
BobN and they share with Alice a Cat-like state
\[
\left| cat\right\rangle _{AB_{1}B_{2}....B_{N}}=\frac{1}{\sqrt{2}}(\left| 0\phi _{1}\phi _{2}....\phi _{N-1}0\right\rangle +\left| 1\phi _{1}^{\prime }\phi _{2}^{\prime }....\phi _{N-1}^{\prime }1\right\rangle )\]
where \( \left| \phi _{i}\right\rangle  \) and \( \left| \phi ^{\prime }_{i}\right\rangle  \)
\( (i=1,2,....,N-1) \) are not necessarily orthogonal. It would then be possible
to make the N Bobs share the state
\[
\left| \chi ^{N}\right\rangle =\alpha \left| \phi _{1}\phi _{2}....\phi _{N-1}0^{\prime }\right\rangle +\beta \left| \phi _{1}^{\prime }\phi _{2}^{\prime }....\phi _{N-1}^{\prime }1^{\prime }\right\rangle \]
initially shared between Claire1, Claire2,....., Claire(N-1) and Alice where
\( \left| \phi _{1}\phi _{2}....\phi _{N-1}0^{\prime }\right\rangle  \) and
\( \left| \phi _{1}^{\prime }\phi _{2}^{\prime }....\phi _{N-1}^{\prime }1^{\prime }\right\rangle  \)
are known but \( \alpha  \), \( \beta  \) are unknown. The particles 1,2,.....,
N belong respectively to Claire1, Claire2,....., Claire(N-1) and Alice. 

Alice first transforms \( \left| \chi ^{N}\right\rangle  \) to 
\[
\left| \xi ^{N}\right\rangle =\alpha \left| \phi _{1}\phi _{2}....\phi _{N-1}0\right\rangle +\beta \left| \phi _{1}^{\prime \prime }\phi _{2}^{\prime \prime }....\phi _{N-1}^{\prime \prime }1\right\rangle \]
where \( \left| \phi _{i}^{\prime \prime }\right\rangle =e^{-i\varepsilon _{i}}\left| \phi _{i}^{\prime }\right\rangle  \),
the \( \varepsilon _{i} \)'s being given by \( \left\langle \phi _{i}\right. \left| \phi _{i}^{\prime }\right\rangle =re^{i\varepsilon _{i}} \)
\( (i=1,2,....,N-1) \). To effect this transformation, Alice has to apply the
unitary operator, on qubit N, that transforms \( \left| 0^{\prime }\right\rangle \rightarrow \left| 0\right\rangle  \)
and \( \left| 1^{\prime }\right\rangle \rightarrow e^{-i\sum ^{N-1}_{i=1}\varepsilon _{i}}\left| 1\right\rangle  \).
Alice also applies the unitary operator, on qubit A, that transforms \( \left| 0\right\rangle \rightarrow \left| 0\right\rangle  \)
and \( \left| 1\right\rangle \rightarrow e^{-i\sum ^{N-1}_{i=1}\varepsilon _{i}}\left| 1\right\rangle  \)
so that \( \left| cat\right\rangle  \) transforms to 
\[
\left| cat_{1}\right\rangle =\frac{1}{\sqrt{2}}(\left| 0\phi _{1}\phi _{2}....\phi _{N-1}0\right\rangle +\left| 1\phi _{1}^{\prime \prime }\phi _{2}^{\prime \prime }....\phi _{N-1}^{\prime \prime }1\right\rangle )\]
The combined state of the 2N+1 particles is now
\[
\left| \xi ^{N}\right\rangle _{12....N}\left| cat_{1}\right\rangle _{AB_{1}B_{2}....B_{N}}\]
which may be written as 
\[
\left( \alpha \left| \phi _{1}\phi _{2}....\phi _{N-1}\phi _{1}\phi _{2}....\phi _{N-1}0\right\rangle +\beta \left| \phi _{1}^{\prime \prime }\phi _{2}^{\prime \prime }....\phi _{N-1}^{\prime \prime }\phi _{1}^{\prime \prime }\phi _{2}^{\prime \prime }....\phi _{N-1}^{\prime \prime }1\right\rangle \right) _{12....(N-1)B_{1}B_{2}....B_{N}}\otimes \left| \phi ^{+}\right\rangle _{NA}\]
\[
\left( \alpha \left| \phi _{1}\phi _{2}....\phi _{N-1}\phi _{1}\phi _{2}....\phi _{N-1}0\right\rangle -\beta \left| \phi _{1}^{\prime \prime }\phi _{2}^{\prime \prime }....\phi _{N-1}^{\prime \prime }\phi _{1}^{\prime \prime }\phi _{2}^{\prime \prime }....\phi _{N-1}^{\prime \prime }1\right\rangle \right) _{12....(N-1)B_{1}B_{2}....B_{N}}\otimes \left| \phi ^{-}\right\rangle _{NA}\]
\[
\left( \alpha \left| \phi _{1}\phi _{2}....\phi _{N-1}\phi _{1}^{\prime \prime }\phi _{2}^{\prime \prime }....\phi _{N-1}^{\prime \prime }0\right\rangle +\beta \left| \phi _{1}^{\prime \prime }\phi _{2}^{\prime \prime }....\phi _{N-1}^{\prime \prime }\phi _{1}\phi _{2}....\phi _{N-1}1\right\rangle \right) _{12....(N-1)B_{1}B_{2}....B_{N}}\otimes \left| \psi ^{+}\right\rangle _{NA}\]
\[
\left( \alpha \left| \phi _{1}\phi _{2}....\phi _{N-1}\phi _{1}^{\prime \prime }\phi _{2}^{\prime \prime }....\phi _{N-1}^{\prime \prime }0\right\rangle -\beta \left| \phi _{1}^{\prime \prime }\phi _{2}^{\prime \prime }....\phi _{N-1}^{\prime \prime }\phi _{1}\phi _{2}....\phi _{N-1}1\right\rangle \right) _{12....(N-1)B_{1}B_{2}....B_{N}}\otimes \left| \psi ^{-}\right\rangle _{NA}\]
As before, Alice now performs a Bell measurement on her two qubits. And Claire\( (i) \)
performs a projection measurement in the orthonormal basis \( \{\left| a_{i}\right\rangle ,\: \left| \overline{a_{i}}\right\rangle \} \)
determined uniquely by 
\[
\left| \phi _{i}\right\rangle =\cos \frac{\theta _{i}}{2}\left| a_{i}\right\rangle +\sin \frac{\theta _{i}}{2}\left| \overline{a_{i}}\right\rangle \]
\[
\left| \phi _{i}^{\prime \prime }\right\rangle =\cos \frac{\theta _{i}}{2}\left| a_{i}\right\rangle -\sin \frac{\theta _{i}}{2}\left| \overline{a_{i}}\right\rangle \]
where \( \theta _{i}\: \epsilon \: [0,\: \pi /2] \) \( (i=1,2,...,(N-1)) \).
Alice and the Claires now communicate their results to the Bobs. By now it is
obvious that whatever be the result at Alice and the Claires, there would always
exist a \emph{product}-unitary operator between the N Bobs so that they (the
Bobs) are left with the state \( \left| \xi ^{N}\right\rangle  \) which may
henceforth be transformed \emph{locally} to \( \left| \chi ^{N}\right\rangle  \).

Let us add that if Alice and the N Bobs share the state
\[
\left| cat^{\prime }\right\rangle =\alpha \left| 0\phi _{1}\phi _{2}....\phi _{N-1}0\right\rangle +\beta \left| 1\phi _{1}^{\prime }\phi _{2}^{\prime }....\phi _{N-1}^{\prime }1\right\rangle \]
the above entanglement teleportation of \( \left| \chi ^{N}\right\rangle  \)
would be possible in a probabilistic manner. 

In all the above cases of entanglement teleportation considered above, the teleported
entangled state is essentially a qubit as each of them is a superposition of
two different states. And in the deterministic cases, Alice shares 1 ebit of
entanglement with the Bobs. Now \emph{1 ebit may be used to teleport at most 1 qubit}.
For if it were possible to teleport 2 qubits, these could \emph{a priori} be
separately entangled maximally with 2 other qubits resulting in the creation
of 2 ebits of entanglement using a single ebit.\cite{10} 

To summarize, we have considered set-dependent entanglement teleportation when
the available channel resource is less than what is needed for universal entanglement
teleportation. In the case of teleportation of entangled states through the
channel \( \frac{1}{\sqrt{2}}(\left| 000\right\rangle +\left| 111\right\rangle )_{AB_{1}B_{2}} \),
the pure entangled states from the plane spanned by known \( \left| 0^{\prime }0^{\prime \prime }\right\rangle  \)
and \( \left| 1^{\prime }1^{\prime \prime }\right\rangle  \) can be exactly
teleported by Alice to Bob1-Bob2 \cite{5}. Note that the amount of entanglement
of these states vary from \( 0 \) to \( 1 \). Shi \emph{et al.}\cite{6} have
considered the probabilistic case when the channel is \( a\left| 000\right\rangle _{AB_{1}B_{2}}+b\left| 111\right\rangle _{AB_{1}B_{2}} \)
and when the states to be teleported are themselves shared between Alice and
a distant party Claire. 

We have shown that an arbitrary pure entanglement from the plane spanned by
known \( \left| \phi 0^{\prime }\right\rangle  \) and \( \left| \phi ^{\prime }1^{\prime }\right\rangle  \)
(where \( \left| \phi \right\rangle  \) and \( \left| \phi ^{\prime }\right\rangle  \)
are arbitrary but fixed, in general, non-orthogonal states) can be deterministically
or probabilistically teleported (from Alice-Claire to Bob1-Bob2) by using the
channel as the GHZ-class states\cite{8} \( \frac{1}{\sqrt{2}}(\left| 0\phi 0\right\rangle +\left| 1\phi ^{\prime }1\right\rangle ) \)
or \( a\left| 0\phi 0\right\rangle +b\left| 1\phi ^{\prime }1\right\rangle  \)
respectively. Note that as the states \( \left| \phi \right\rangle  \) and
\( \left| \phi ^{\prime }\right\rangle  \) are in general non-orthogonal states,
the amount of the entanglement of the teleported states vary from \( 0 \) to
\( e \), where \( e\leq 1 \). And our protocol is deterministic or probabilistic
depending on whether Alice shares one ebit with the Bob1-Bob2 system or less
than that. It is interesting to note that although for both \( \left| GHZ\right\rangle =\frac{1}{\sqrt{2}}(\left| 000\right\rangle +\left| 111\right\rangle )_{AB_{1}B_{2}} \)
and \( \left| ghz\right\rangle =\frac{1}{\sqrt{2}}(\left| 0\phi 0\right\rangle +\left| 1\phi ^{\prime }1\right\rangle )_{AB_{1}B_{2}} \)
Alice shares 1 ebit with the Bobs, the channel between Alice and Bob2 is distillable
for \( \left| ghz\right\rangle  \) (when \( \left| \phi \right\rangle  \)
and \( \left| \phi ^{\prime }\right\rangle  \) are non-orthogonal) and separable
for \( \left| GHZ\right\rangle  \) while the channels between Alice and Bob1
are separable for both.\cite{12} This could somehow be the reason as to why
\( 0 \) to \( 1 \) entanglement is \emph{not} transferred through \( \left| ghz\right\rangle  \)
(when \( \left| \phi \right\rangle  \) and \( \left| \phi ^{\prime }\right\rangle  \)
are non-orthogonal) although the same is possible through \( \left| GHZ\right\rangle  \).
We then generalized these considerations to the case in which the state to be
teleported is an N-party entangled state. 

U. S. thanks Dipankar Home for encouragement and acknowledges partial support
by the Council of Scientific and Industrial Research, Government of India, New
Delhi.


\begin{thebibliography}{}
\bibitem{7}So earlier operations performed by Alice are not possible now.
\bibitem{10}Here we have used the fact that in teleportation of a single qubit A (possessed
by say, Alice) to a possibly distant party say, Bob, through an entangled channel
between Alice and Bob, an initial entanglement between A and C (possessed by
say, Claire, possibly distant from both Alice and Bob) is fully transferred
between C and Bob's particle. See \cite{11}
\bibitem{12}Here by ``channel between Alice and Bob1'', we mean that part of the channel
which is left when Bob2 is traced out. Similarly for ``channel between Alice
and Bob2''.
\end{thebibliography}
\end{document}